\documentclass[12pt,epsfig]{article}
\usepackage{epsfig}
\newcommand{\singlespace}{
    \renewcommand{\baselinestretch}{1}\large\normalsize}

\begin{document}

\begin{titlepage}
\pagestyle{empty}
\setcounter{page}{1}
\vspace{1.0in}
\begin{center}
\singlespace
\begin{LARGE}
{\bf Pseudoscalar neutral mesons in hot and dense matter}\\
\end{LARGE}
\end{center}
\vskip 1.0in
\begin{center}
{\large {\bf P. Costa}}\footnote{pcosta@teor.fis.uc.pt}, 
{\large {\bf M. C. Ruivo}}\footnote{maria@teor.fis.uc.pt}

{\em  Departamento de F\'{\i}sica
da Universidade de Coimbra

P - 3004 - 516 Coimbra, Portugal}

\vspace{0.3cm}

{\large {\bf Yu. L. Kalinovsky}}\footnote{kalinov@nusum.jinr.ru}

{\em  Laboratory of Information Technologies,

Joint Institute for Nuclear Research, Dubna, Russia}
\end{center}
\vspace{2cm}

\begin{abstract}
The behavior of neutral pseudoscalar mesons $\pi^0\,, \eta$ and $\eta'$ in hot and dense matter  
is investigated, in the framework of the three flavor Nambu-Jona-Lasinio model. Three different 
scenarios are considered: zero density and finite temperature,  zero temperature and finite density 
in a flavor asymmetric medium with and without strange valence quarks, and finite 
temperature and density. The behavior of mesons is analyzed in connection with possible signatures 
of restoration of symmetries. In the high density region and  at zero temperature it is found that 
the mass of the $\eta'$ increases, the deviation from the mass of the $\eta$ being more pronounced in   
matter without strange valence quarks. 

\end{abstract}
\end{titlepage}
\newpage
\setcounter{page}{2}

\section{Introduction}
\hspace*{\parindent}Different regions of the QCD phase diagram are object of interest nowadays. 
Understanding matter under extreme conditions is relevant to understand the physics of heavy-ion collisions, 
the early universe and neutron stars. The dominant degrees of freedom of QCD at high T and $\mu$ 
are supposed to be partons (a quark-gluon plasma) instead of hadrons and  restoration of 
symmetries is expected to occur.  While  the phase transition at zero chemical potential and finite 
temperature is accepted to be second order or crossover, there  are indications that the
phase transition with finite chemical potential and zero temperature is first order \cite{kanaya}.  
Experimental and theoretical efforts have been done  in order to explore the $\mu-T$ phase boundary. 
Recent Lattice results indicate a critical "endpoint", connecting the  first order phase transition 
with the crossover region at $T_E=160\pm35\mbox{MeV}\,,\mu_E=725\pm35\mbox{MeV}$ \cite{Fodor}. 
Understanding the results of experiments at BNL \cite{rhic} and  CERN \cite{cern} provides a 
natural  motivation for these studies.  

Much attention has  been paid to the  question of which symmetries are restored.
In the limit of vanishing quark masses the QCD Lagrangian has 8 Goldstone bosons,
associated with the dynamical breaking of chiral symmetry.  In order to give a finite
mass to the mesons, chiral symmetry is broken {\it ab initio} by giving current masses
to the quarks.The mystery of the non-existence of the ninth Goldstone boson,
predicted by the quark model, was solved by assuming that the QCD Lagragian
has a $U_A (1)$ anomaly. Explicitly breaking the $U_A (1)$ symmetry,
for instance by instantons, has the  effect of giving a mass
to $\eta'$ of about $1$ GeV. So the mass of $\eta'$ has a different origin
than the masses of the pseudoscalar mesons and this meson cannot be regarded
as the remanent of a Goldstone boson.

The study of the behavior of mesons  in hot and dense matter is
an important issue since they might provide a signature of the phase
transition and give indications about which symmetries are restored.
According to Shuryak \cite{shuryak}, there are two scenarios: only $SU(3)$ chiral
symmetry is restored or both, $SU(3)$ and $U_A(1)$, symmetries are restored.
The behavior of $\eta'$ in medium or of related observables, like the topological
susceptibility \cite{ohta1} might help to decide between these scenarios.
A decrease of the $\eta'$ mass in medium could manifest in the
increase of the $\eta'$ production cross section, as compared to that for $pp$ 
collisions \cite{kapusta}.

Theoretical concepts  on the behavior of matter at high densities have been
mainly developed from  model calculations. The
Nambu-Jona-Lasinio [NJL] type models \cite{njl} have been extensively used
over the past years to describe low energy features of hadrons and also to
investigate restoration of chiral symmetry with temperature or density
\cite{njl,kunihiro,njlT,weise,maria,RuivoSousa,RSP,buballa,costaruivo,klev2}.

The NJL model is an effective quark model that, besides its simplicity, has the advantage
of incorporating important symmetries of QCD and of exhibiting also the symmetry breaking mechanism. 
Since the model has no confinement,  several drawbacks are well known. For instance, the $\eta'$ 
mass lies above the $\bar q q$ threshold.  The model describes this meson as $\bar q q$ resonance, 
which would have the unphysical decay in $\bar q q$ pairs,  and  the definition of its mass is
unsatisfactory. For this reason,  it has been proposed that in order to
investigate the possible restoration of $U_A(1)$ symmetry, instead of $m_{\eta'}$,
the topological susceptibility $\chi$, a more reliable quantity in this model, should be used \cite{ohta1}.

The behavior of $SU(3)$ pseudoscalar mesons, in NJL models,  with temperature has
been studied in \cite{kunihiro,njlT,klev2}. Different studies have been devoted to the
behavior of pions and kaons at finite density in flavor symmetric
or asymmetric matter \cite{maria,RuivoSousa,RSP,costaruivo}.

This paper is devoted to investigate the phase transition in hot and dense
flavor asymmetric matter and the  in medium behavior of the neutral $SU(3)$ pseudoscalar mesons,
in the framework of the $SU(3)$ NJL model including the 't Hooft
interaction, which breaks the $U_A(1)$ symmetry.  We choose these mesons because they might carry 
information about the restoration of symmetries: pions, which have been studied extensively 
in this concern, are a privileged tool to appreciate the restoration of chiral symmetry, since, 
being the lightest pseudoscalar mesons, are the best Goldstone-like bosons. On the other side, 
the in-medium behavior of  $\eta\, \mbox{and}\, \eta'$ is expected to carry information about 
the restoration of $U_A(1)$ symmetry.  At variance with
the other mesons, there is not much information about the behavior of these mesons in medium.
Concerning the $\eta$ meson, the recent discovery of mesic atoms might provide
useful information in this concern \cite{sokol}.


\section{Model and Formalism}\label{model}

\hspace*{\parindent}We use a version of the $SU(3)$ NJL model described by  the Lagrangian:
\begin{eqnarray}
{   L} &=& \bar{q} \left( i \partial \cdot \gamma - \hat{m} \right) q
+ \frac{g_S}{2} \sum_{a=0}^{8}
\Bigl[ \left( \bar{q} \lambda^a q \right)^2+
\left( \bar{q} (i \gamma_5)\lambda^a q \right)^2
 \Bigr] \nonumber \\
&+& g_D \Bigl[ \mbox{det}\bigl[ \bar{q} (1-\gamma_5) q \bigr]
  +  \mbox{det}\bigl[ \bar{q} (1-\gamma_5) q \bigr]\Bigr] \, .
\label{lagr}
\end{eqnarray}
Here $q = (u,d,s)$ is the quark field with three flavors, $N_f=3$, and
three colors, $N_c=3$. $\lambda^a$ are the Gell - Mann matrices, a = $0,1,\ldots , 8$,
${ \lambda^0=\sqrt{\frac{2}{3}} \, {\bf I}}$.
The explicit symmetry breaking part (\ref{lagr})
contains the current quark masses $\hat{m}=\mbox{diag}(m_u,m_d,m_s)$.
The last term in (\ref{lagr}) is the lowest six-quark dimensional operator and it
 has the $SU_L(3)\otimes SU_R(3)$ invariance but breaks the
$U_A(1)$ symmetry. This term is a reflection of the axial anomaly in QCD.
For the general reviews on the three flavor version of the NJL model
see Refs. in \cite{njl,kunihiro,njlT}.

To obtain a four quark interaction  from the six quark interaction we make a shift
$(\bar{q} \lambda^a q) \longrightarrow (\bar{q} \lambda^a q) + <\bar{q} \lambda^a q>$,
where $<\bar{q} \lambda^a q>$ is the vacuum expectation value, and
contract one bilinear $(\bar{q} \lambda^a q)$.  The hadronization procedure can be done 
by the integration over quark fields in the functional integral, 
leading to  the following effective action: 

\begin{eqnarray}\label{act}
  W_{eff}[\varphi,\sigma] &=&
  - \frac{1}{2} \left( \sigma^a S^{-1}_{ab}\sigma^b \right)
  - \frac{1}{2} \left( \varphi^a P^{-1}_{ab}\varphi^b \right)
  \nonumber \\ &&
  -i \mbox{Tr} \,  \mbox{ln} \Bigl[ i (\gamma_\mu \partial_\mu ) -
  \hat{m} + \sigma_a \lambda^a
  + (i \gamma_5 )(\varphi_a \lambda^a) \Bigr] \, .
\end{eqnarray}

The fields $\sigma^a$ and $\varphi^a$ are scalar and pseudoscalar meson nonets. 
We have introduced projectors
\begin{eqnarray}
  S_{ab} &=& g_S \delta_{ab} + g_D D_{abc}<\bar{q} \lambda^c q>, \label{sab}\\
  P_{ab} &=& g_S \delta_{ab} - g_D D_{abc}<\bar{q} \lambda^c q>. \label{pab}
\end{eqnarray}

 The constants $D_{abc}$ are such that they 
coincide with the $SU(3)$ structure constants $d_{abc}$ for $a,b,c =(1,2,\ldots ,8)$
and $D_{0ab}=-\frac{1}{\sqrt{6}}\delta_{ab}$, $D_{000}=\sqrt{\frac{2}{3}}$.

A straightforward generalization of the model for finite temperature and density  can be 
done by using the  Matsubara technique (see \cite{klev2,crk}). The first variation of the 
action (\ref{act}) leads to the  well known gap equations. By expanding the effective action
(\ref{act}) over meson fields and keeping the pseudoscalar mesons only,
we have the effective  action:
\begin{eqnarray}\label{act2}
  W_{eff}^{(2)}[\varphi] =
  -\frac{1}{2}\varphi^a \left[ P_{ab}^{-1} - \Pi_{ab} (P) \right] \varphi^b
= -\frac{1}{2}\varphi^a  D_{ab}^{-1}(P)  \varphi^b
\end{eqnarray}
where $D_{ab}^{-1}(P)$ is the inverse unnormalized meson propagator and $\Pi_{ab}(P)$ is  
the polarization operator, which in the momentum space has the form
\begin{eqnarray}\label{polop}
\Pi_{ab} (P) = i N_c \int \frac{d^4p}{(2\pi)^4}\mbox{tr}_{D}\left[
S_i (p) (\lambda^a)_{ij} (i \gamma_5 )
S_j (p+P)(\lambda^b)_{ji} (i \gamma_5 )
\right] \, ,
\end{eqnarray}

The model is fixed by coupling constants $g_S, g_D$, the cutoff parameter $\Lambda$
which regularizes momentum space integrals and current quark masses.
We use the parameter set 
$m_u = m_d = 5.5$ MeV,  $m_s = 140.7$ MeV, $g_S \Lambda^2 =  3.67\,, g_D \Lambda^5 = -12.36$ 
and $\Lambda =602.3$ MeV, that has been determined by fixing the conditions: 
$M_{\pi^0} = 135.0$ MeV, $M_K   = 497.7$ MeV, $f_\pi =  92.4$ MeV and $M_{\eta'}= 960.8$ MeV. 
We also have: 
$M_{\eta}= 514.8$ MeV, $\theta (M_{\eta}^2) = -5.8^{\circ}$, $ g_{\eta \bar{u}u} =   2.29$,
$g_{\eta \bar{s}s} = -3.71$ 
$M_{\eta'}= 960.8$ MeV, $\theta (M_{\eta'}^2) = -43.6^{\circ}$, $ g_{\eta'\bar{u}u} =  13.4$,
$g_{\eta' \bar{s}s} = -6.72$.

Although, due to the lack of confinement in the model, the $\eta'$ - meson is here described  
as a resonance state of $\bar q q$,  we use  $M_{\eta'}$ as an input parameter.

To consider the diagonal  mesons $\pi^0$, $\eta$ and $\eta'$ we  take into account
the matrix structure of the propagator in (\ref{act2})  
(See Ref. \cite{kunihiro,klev2,crk} for details). 
In the basis of $\pi^0 - \eta - \eta'$ system we write the projector $P_{ab}$ and the 
polarization operator   $P_{ab}$ as  matrices:
\begin{eqnarray}
&&  {P}_{ab} =
  \left(
\begin{array}{ccc}
P_{33} & P_{30} & P_{38} \\
P_{03} & P_{00}& P_{08} \\
P_{83} & P_{80}& P_{88}
\end{array}
\right)\,\,\,\,\,\,\mbox{and}\,\,\,\,\,\,{\Pi}_{ab} =
  \left(
\begin{array}{ccc}
\Pi_{33} & \Pi_{30} & \Pi_{38} \\
\Pi_{03} & \Pi_{00}& \Pi_{08} \\
\Pi_{83} & \Pi_{80}& \Pi_{88}
\end{array}
  \right).
\end{eqnarray}
It should be noticed that in media with different densities of $u\, \mbox{and}\,d$ 
quarks the nondiagonal  matrix elements $ P_{30}\,, P_{38}\propto(<<\bar{q}_uq_u>>-<<\bar{q}_dq_d>>)$
and $\Pi_{30}\,, \Pi_{38}\propto(I_2^{uu}(P) -I_2^{dd}(P))$  are non vanishing 
and correspond to  $\pi^0 - \eta$ and $\pi^0 -\eta'$ mixing (here $<<\bar{q_i}q_i>>$ 
are the in medium quark condensates and $I_2^{ii}(P)$ are write below). Here we did the 
approximation of neglecting these elements, whose effects we found to be negligible by 
means of a simple  estimation. The nonvanisnhing elements of $\Pi_{ab}$ depend on the 
regularized integrals \cite{crk}:
\begin{eqnarray}\label{sint}
I_2^{ii}(P_0, T, \mu_i) =
&& - \frac{N_c}{2\pi^2} {\mathcal{P}} \int \frac{{\tt p}^2 d {\tt p}}{E_i} \,\,
\frac{1}{P_0^2-4 E_i^2} \left( n^+_i - n^-_i\right)
\nonumber \\
&& - i  \frac{N_c}{4\pi} \sqrt{ 1- \frac{4 M_i^2}{P_0^2}  }
\left(n^+_i(\frac{P_0}{2}) - n^-_i (\frac{P_0}{2})\right) \,,
\end{eqnarray}

where $n_i^{\mp}= n_i^{\mp}(E_i)$ 
are the Fermi distribution functions of the negative (positive) energy state
of the $i$th quark. Notice that, as the temperature and/or density increase these integrals may 
acquire an imaginary part, as long as the meson masses cross the $\bar{q} q $ threshold. 
This is also true for $\eta'$ in the vacuum, as already mentioned.  


\section{Discussion and conclusions}
\hspace*{\parindent}We present our results for the nature of the phase transition and  for the masses of 
$\pi^0\,,\eta\,\mbox{and}\,\eta'$ in three scenarios: zero density and finite temperature,  
zero temperature and finite density and finite temperature and density.  Two types 
of asymmetric quark matter that are considered: for the  first  case (Case I), that might be formed 
temporarily in  heavy-ion collisions,  we fix densities by $\rho_d=2\rho_u$ and $\rho_s=0$; 
for the second one (Case II), that might exist in neutron stars, we impose the condition of 
$\beta$ equilibrium  and charge 
neutrality through the following constraints, respectively on the chemical potentials and densities 
of quarks and electrons $\mu_{d}=\mu_{s}=\mu_{u}+\mu_{e}\,\, \mbox{ and }\,\,\,\frac{2}{3}\rho_{u}-\frac{1}{3}(\rho_{d}+\rho_{s})-\rho_{e}=0$, with 
$\rho_{i}=\frac{1}{\pi^{2}}(\mu_{i}^{2}-M_{i}^{2})^{3/2}\theta(%
\mu_{i}^{2}-M_{i}^{2})\,\,\mbox{ and }\,\,\,\rho_{e}=\frac{\mu_{e}^{3}}{3\pi^{2}}.$

\begin{figure}[t]
\begin{center}
\epsfig{file=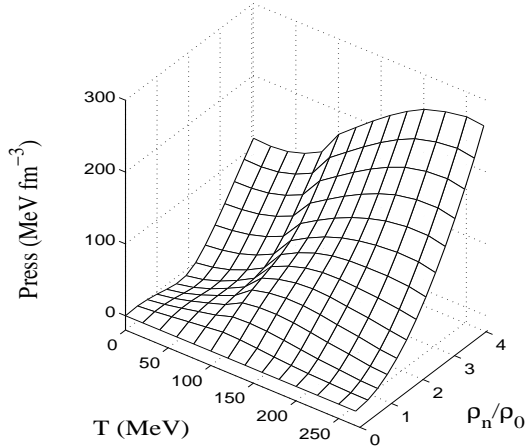, width=7cm,height=6cm}
\end{center}
\caption{Pression as function of density and temperature.}
\end{figure}

The nature of the phase transition in NJL models at finite T and/or $\mu$ has been discussed by 
different authors \cite{njlT,weise,RSP,buballa,costaruivo}. At zero density the phase transition 
is a smooth crossover one; at non zero densities different situations occur. Here we analyze this 
question  by examining  the curve of the pressure as a function of density and temperature (Fig. 1), 
defined as $P(\rho, T) = - [ \Omega(\rho, T) - \Omega(0, T) ]$,
where $\Omega(\rho, T)$ is the thermodynamic potential \cite{weise,RSP,greiner}. At zero temperature 
and finite density,  as it has been shown in \cite{RSP,buballa}, there is a region of densities 
where the pressure is negative, meaning that a phase of low density and broken symmetry coexists 
with a phase of high density and partially restored chiral symmetry (see Fig. 1). 
A suitable choice of parameters leads to the first zero of the pressure at $\rho\sim 0$ 
and the second zero,   (at $\rho_c=2.25$, for  Case I and Case II ) corresponds to a minimum 
of the energy per particle, which means that  a stable a hadronic phase exists at these densities, 
consisting of droplets of quarks surrounded by a non trivial vacuum. 
As pointed out in \cite{costaruivo} different features occur in the quark phase 
( for $\rho > \rho_c$) according to we consider or not matter in beta equilibrium.  
In the last case   the mass of the strange quark  becomes     lower than the chemical potential  
at densities above $\sim 3.8\rho_0$ what implies the occurrence   of strange quarks in this regime. 
As it will be discussed below, this fact leads to meaningful differences in the behavior of $\eta\,,\eta'$.

\begin{figure}[t]
\begin{center}
\hspace*{-0.2cm}\epsfig{file=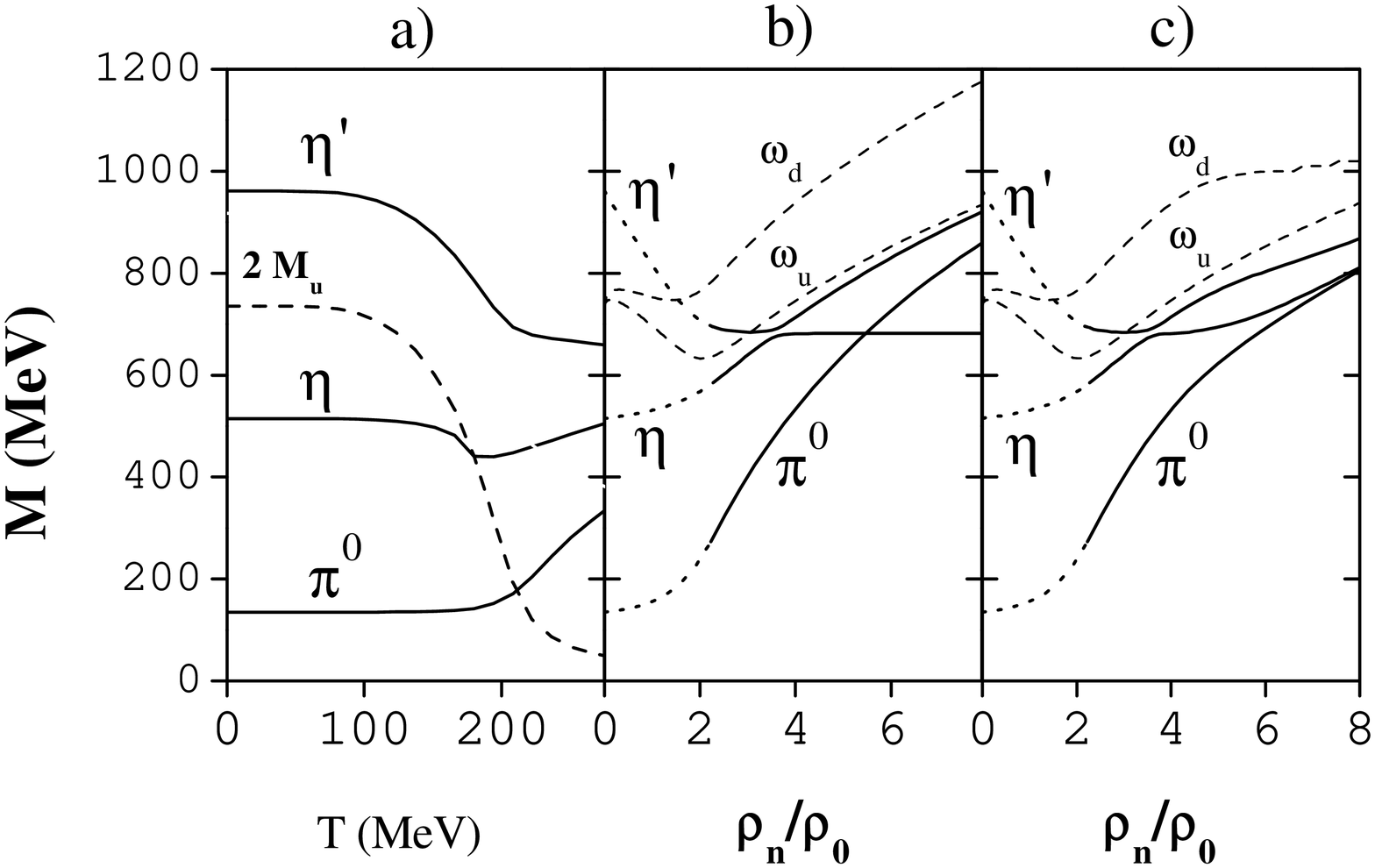, width=12.5cm,height=7cm}
\end{center}
\caption{
$\eta$, $\eta'$ and $\pi^0$ masses as function of temperature a)
and density: b) Case I; c) Case II.
}
\end{figure}

Now we discuss the nature of the phase transition at finite density and temperature by studying 
the pressure in the $T-\rho$ that is shown in Fig. 1. One can see that there is a region of the 
surface with negative curvature that corresponds to the region of temperatures and densities where 
the phase transition is first order, since  the pressure and/or compressibility are negative. 
The point $T= 56\mbox{Mev}, \rho=1.53 \rho$, were the pressure is already positive and the 
compressibility has only one zero, is identified, as usual, as  the critical endpoint,  that 
connects the first order and the second order phase transition regions. Above that point, 
we have a smooth phase transition, corresponding to the regions of positive curvature of the surface. 

Unlike at $T=0$,  we can not talk anymore about a stable hadronic phase in the first order phase 
transition region  because the absolute minimum of the energy per particle is now at zero density. 
In spite of these drawbacks, we think it is illustrative to plot  the masses of 
$\pi^0\,, \eta\,$ and $\eta'$ in the whole region of the $\rho-T$ plane. 

Let  us now analyze the in medium behavior of the mesons under study. To understand our  results, 
it is useful to study the limits of the Dirac sea (denoted by $\omega_i$) of this mesons. 
They can be obtained by looking the limits of the regions  of poles in  the integrals $I_2^{ii} (P_0)$ 
(see (\ref{sint})): $\omega_{u}= 2\mu_{u}$, 
$\omega_{d}= 2\mu_{d}$ and $\omega_{s}= 2\mu_{s}$. (At finite temperature, we will have 
$\omega_{u}= 2M_u$, $\omega_{d}= 2M_d$ and $\omega_{s}= 2M_s$).

The behavior of these mesons at $\rho = 0$ and $T\not=0$ has been study by  \cite{kunihiro,klev2}. 
We reproduce these results for the sake of comparison with our new findings at $T=0\,,\rho\not=0$ 
(see Fig. 2 a)). The mesonic  masses   change smoothly and when they exceed the sum of the masses 
of the constituent quarks the mesons became unbound (Mott temperature). Usually,  the critical 
temperature is defined as the Mott transition temperature for the pion. The masses of 
$\eta\,\mbox{and}\,\eta'$ show a tendency to become degenerated, even after the critical temperature. 
The decrease of the mass of $\eta'$ is not generally considered  enough to give an indication 
of restoration of $U_A(1)$ symmetry in hot media \cite{ohta1,kapusta}.

\begin{figure}[t]
\begin{center}
\epsfig{file=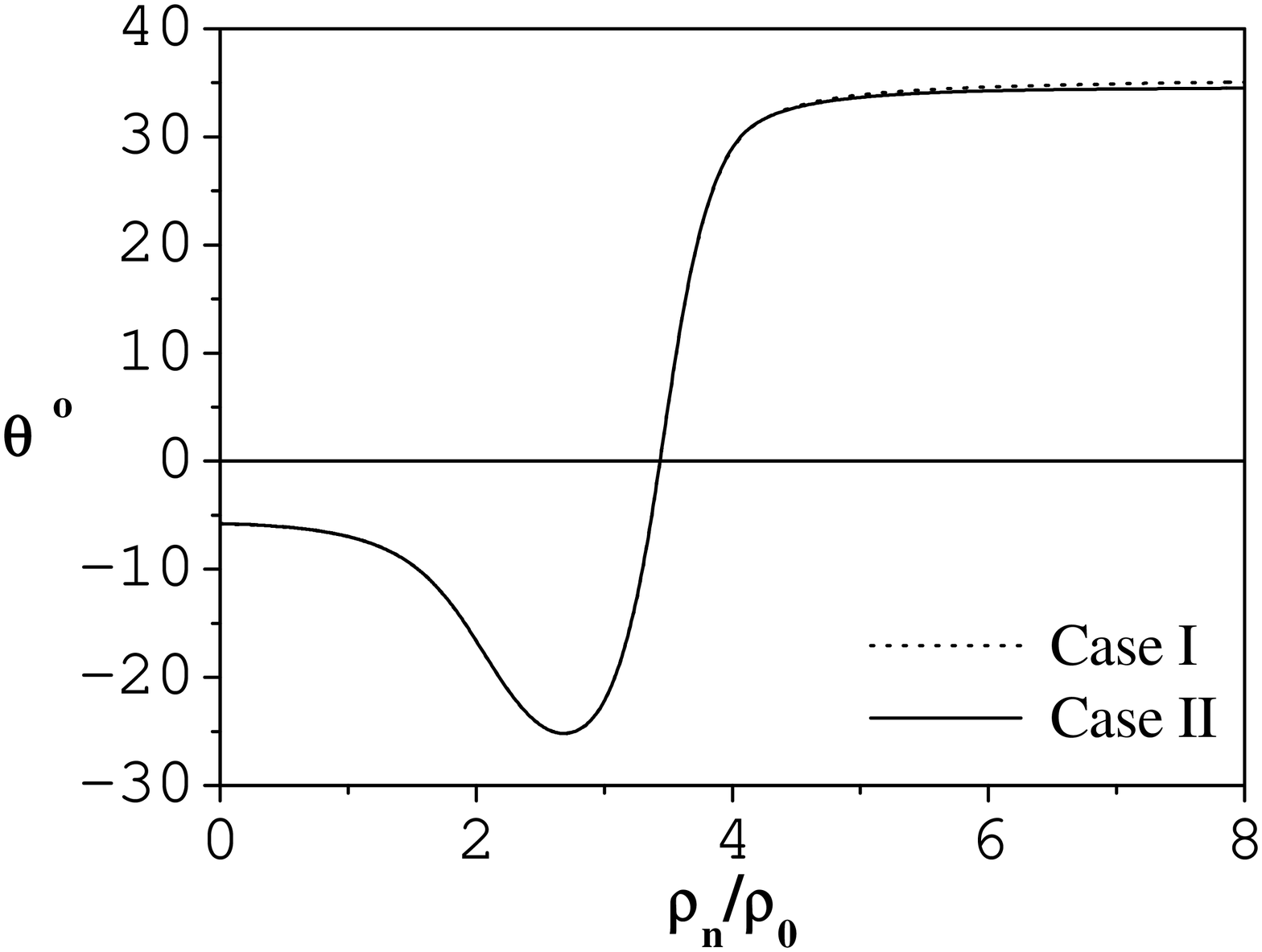, width=8cm,height=6cm}
\end{center}
\caption{Mixing angle, $\theta$, as function of density.}
\end{figure}

Now, we analyze the results  for the masses of $\eta$ and $\eta'$ at finite densities and zero 
temperature, that are plotted in Fig. 2 b)-c). Up to the critical density, they exhibit a tendency 
to became degenerated, but, after that point they split again, the splitting being more 
pronounced in Case I. These results show that if a degeneracy  of $\eta\,,\eta'$ is an indication 
of the restoration of $U_A(1)$ this is unlikely at high densities. In Case II  the system shows a 
tendency to the restoration of  flavor symmetry, which is related with the presence of 
strange quarks in the medium that occurs at about $\rho\sim 3.8 \rho_0$ and, as shown in 
\cite{costaruivo}, manifests itself in other observables.  As we can see, the $\eta'$ - meson lies 
above the quark - antiquark threshold for $\rho_n < 2.5 \rho_0$ and it is a resonant state. 
After that density, the $\eta'$ becomes a bound state. We checked that the behavior in matter
with $\rho_u=\rho_d,\rho_s=0$ is qualitatively similar to Case I. 
 
In order to understand the difference between the results  in the high density region for 
Cases I and II, it is convenient to study the in medium behavior of the mixing angle, $\theta$, 
that is plotted in Fig. 3. As it is well known, the quark content of $\eta$ and $\eta'$ depend 
on the mixing angle in the form:
\begin{eqnarray}\label{mix}
|\eta> &=& \cos \theta \frac{1}{\sqrt{6}} |\bar u u + \bar d d - 2 \bar s s> - \sin \theta \frac{1}{\sqrt{3}} 
|\bar u u + \bar d d + \bar s s>,\\
|\eta'> &=& \cos \theta \frac{1}{\sqrt{3}} |\bar u u + \bar d d + \bar s s> + \sin \theta \frac{1}{\sqrt{6}} 
|\bar u u + \bar d d - 2 \bar s s>.
     \end{eqnarray}

Above $\rho\simeq 3.5 \rho_0$ the angle becomes positive and increases rapidly and the 
strange quark content, $y = (\bar u u + \bar d d )/ \bar s s$,  of the mesons changes: 
at low density, the $\eta '$ is more strange than  the $\eta$ but the opposite occurs at high density. 
Since in Case I there are no strange quarks in the medium (what implies that the strange quark mass 
is almost unaffected), the $\eta$ mass should stay constant in the region of densities where its 
content is dominated by the strange quark. This explains the larger splitting between 
$\eta$ and $\eta '$ in Case I.

\newpage
\begin{figure}
\begin{center}
  \begin{tabular}{cc}
    \hspace*{-0.2cm}\epsfig{file=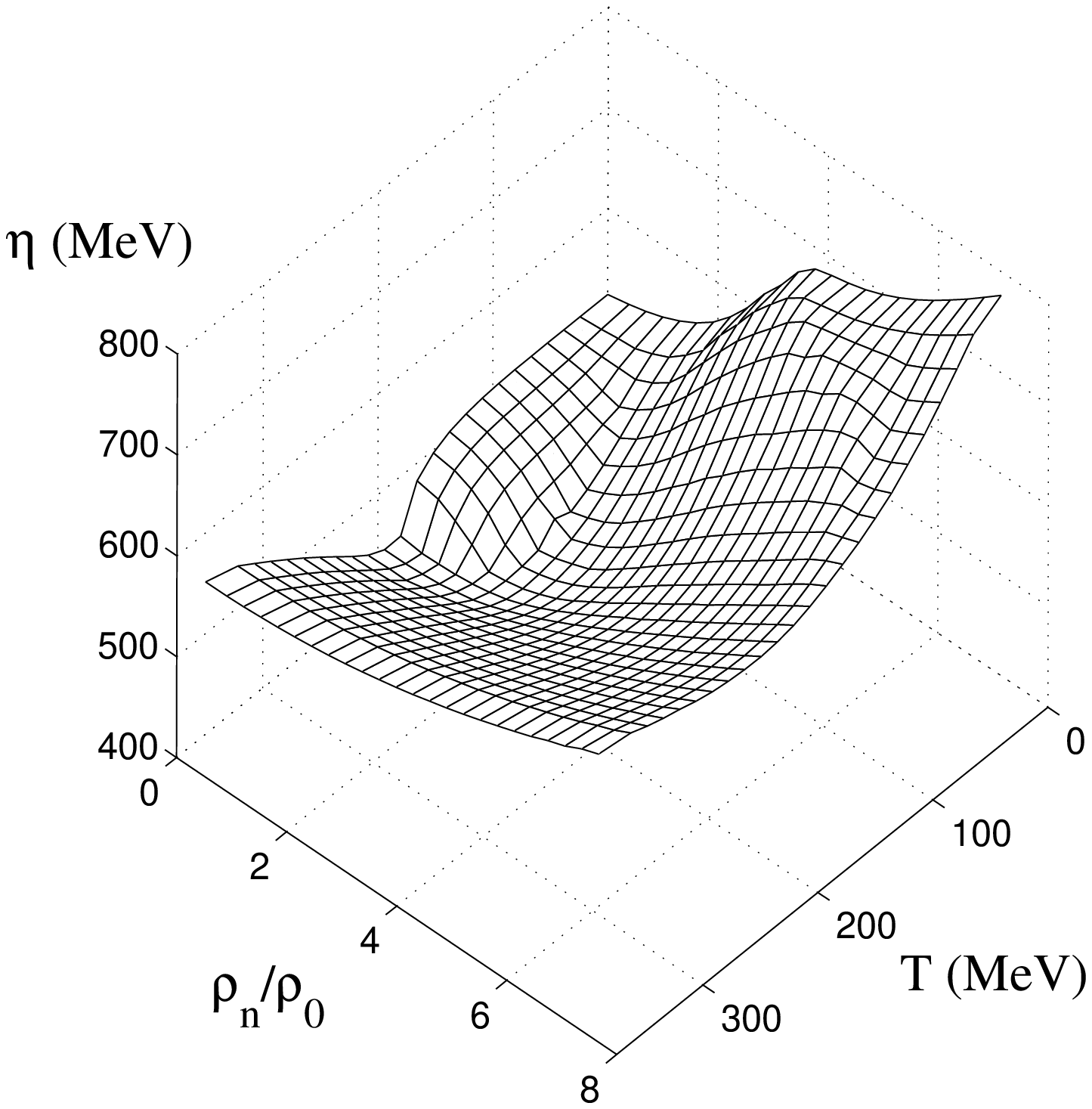,width=6.5cm,height=6.5cm} &
    \epsfig{file=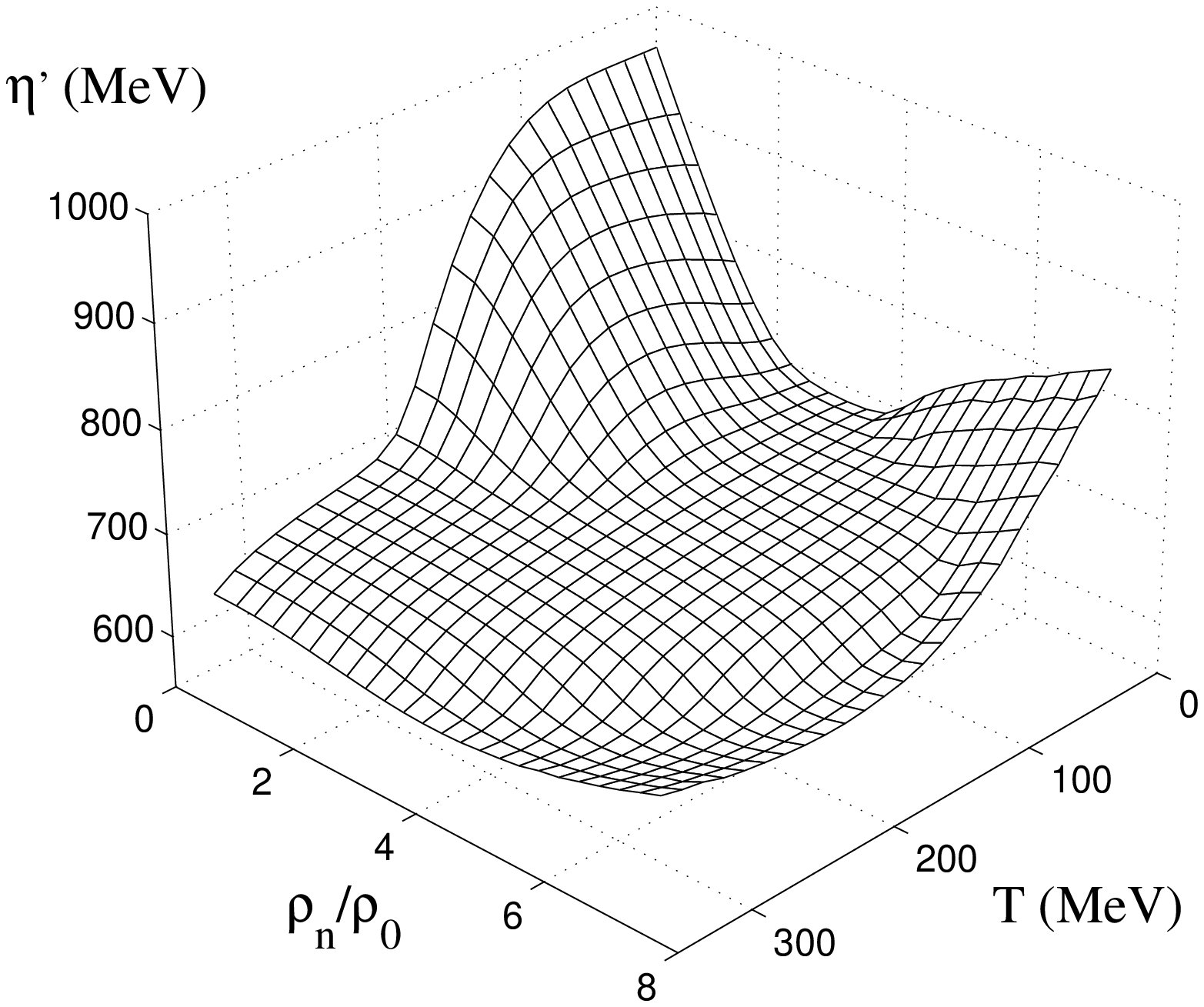,width=6.5cm,height=6.5cm} \\
   \end{tabular}
   \begin{center}
    	\epsfig{file=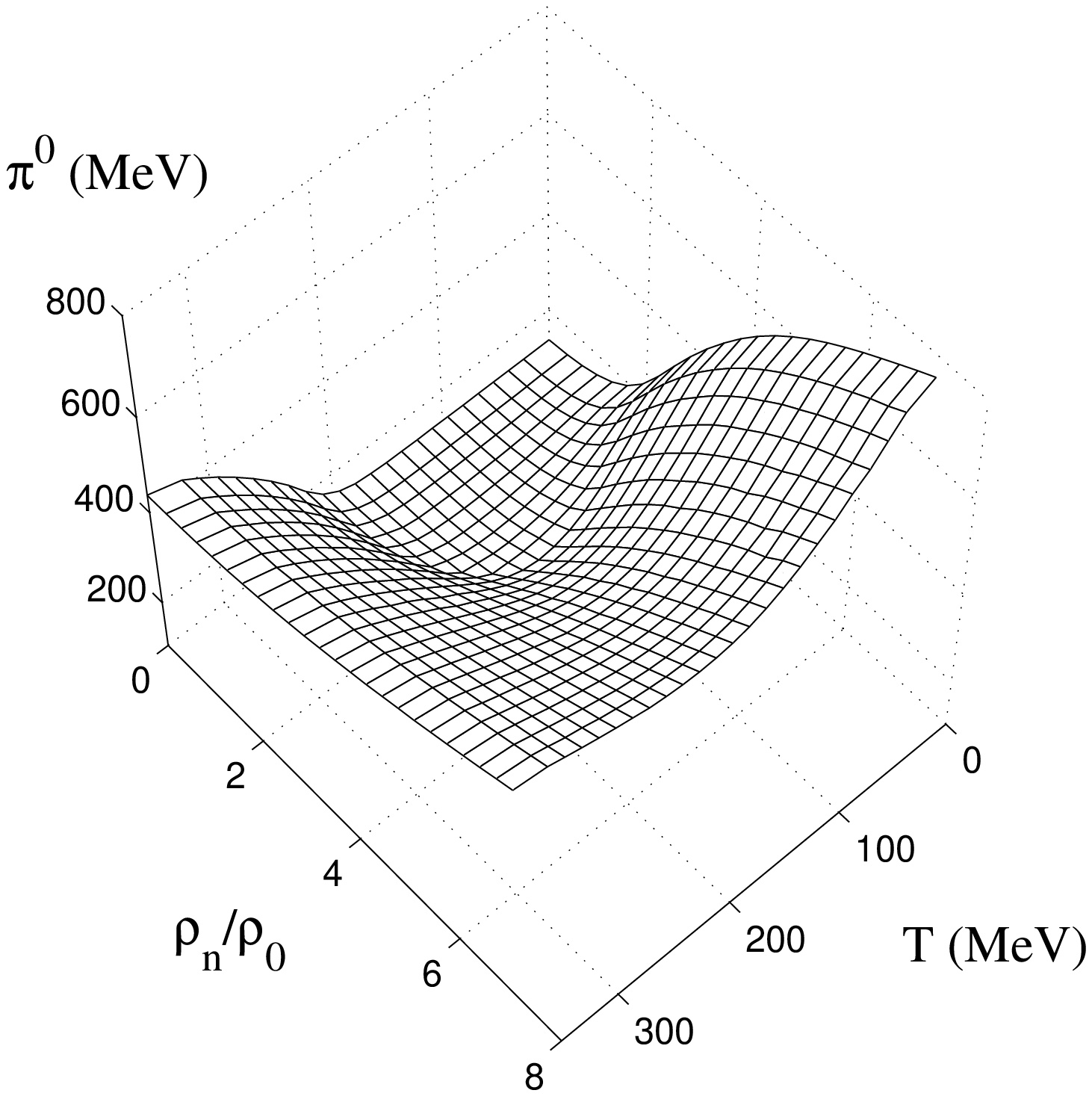,width=6.5cm,height=6.5cm} 
   \end{center}
\end{center} 
\caption{
Mesons masses as functions of temperature and density (Case II).
Upper panel: $\eta$ and $\eta'$ mesons.
Lower panel: $\pi^0$ meson.
}
\end{figure}

Considering now the  finite density and temperature case (we plot the masses of the
mesons for the case of quark matter with strange quarks in $\beta$ equilibrium at finite 
$T$ in Fig. 4),  and  similarly to the finite temperature and zero density case, 
the $\bar q q$ threshold for the different mesons is again at  the sum of the constituent 
quark masses, so the mesons dissociate at densities and temperatures close to the critical ones.  
A second feature to be noticed is that the  mesons that have a Goldstone boson like  nature 
show more clearly the difference between the chiral symmetric and asymmetric phase.  
The diagram for $\pi^0$ shows clearly a "line" that separates the chiral broken phase 
(region of negative curvature) from the chiral restored phase. This is not so evident for the 
$\eta$. A more smooth  behavior is found for $\eta'$ (Fig. 4), which is natural, 
since it is not a Goldstone boson associated with chiral symmetry.   

\newpage
\begin{center}
{\large Acknowledgment:}
\end{center}
Work supported by grant SFRH/BD/3296/2000 (P. Costa), Centro de F\'{\i}sica Te\'orica, 
FCT and GTAE (Yu. L. Kalinovsky).



\end{document}